\documentclass[prl,reprint,superscriptaddress]{revtex4-1}

\usepackage{amsmath} 
\usepackage{amssymb}  
\usepackage{amsfonts} 
\usepackage{dsfont}
\usepackage{bm}  
\usepackage{bbm}   
\usepackage{bbold}   
\usepackage{braket} 
\usepackage{color} 
\usepackage{comment}  
\usepackage{dcolumn}  
\usepackage{enumerate}  
\usepackage{epsfig}  
\usepackage{gensymb}  
\usepackage{graphicx}  
\usepackage{indentfirst}  \usepackage{lmodern}  
\usepackage{mathrsfs}  
\usepackage{mathtools}  
\usepackage{psfrag}  
\usepackage{pst-all} 
\usepackage{soul}  
\usepackage{xcolor}
\usepackage{float} 
\usepackage[colorlinks,linkcolor=blue,citecolor=blue,urlcolor=blue,hyperindex,driverfallback=dvipdfm]{hyperref}  \usepackage[T1]{fontenc} 

\usepackage{dirtytalk}
\usepackage{CJKutf8}

\usepackage{float} 
\usepackage{placeins}
\usepackage{bm}
\usepackage{enumitem}

 \usepackage[normalem]{ulem} 
 \makeatletter
\newcommand*{\addFileDependency}[1]{
  \typeout{(#1)}
  \@addtofilelist{#1}
  \IfFileExists{#1}{}{\typeout{No file #1.}}
}
\makeatother
 
\newcommand*{\myexternaldocument}[1]{%
    \externaldocument{#1}%
    \addFileDependency{#1.tex}%
    \addFileDependency{#1.aux}%
}
\usepackage[strict]{changepage}
\usepackage{calc}
\usepackage{dirtytalk}
\usepackage{pifont}

\usepackage{xr-hyper}

\usepackage{graphicx}
\usepackage{cancel}
\usepackage{longtable}
\usepackage{array}
\newcolumntype{C}[1]{>{\centering\arraybackslash}p{#1}}

\usepackage{ulem}
\usepackage{titlesec}
 \titleformat{\paragraph}[hang]{\bfseries}{}{0pt}{\uline}
\titlespacing*{\paragraph}
{0pt}{3.25ex plus 1ex minus .2ex}{1.5ex plus .2ex}

\usepackage{dirtytalk}
\usepackage{CJKutf8}
\usepackage{dblfloatfix}
\myexternaldocument{supplementary}

\begin{document}
\begin{CJK*}{UTF8}{gbsn}
\title{ Strong field physics in open quantum systems}
\author{Neda Boroumand }
\affiliation{Department of Physics, University of Ottawa, Ottawa, Ontario K1N 6N5, Canada}
\author{Adam Thorpe}
\affiliation{Department of Physics, University of Ottawa, Ottawa, Ontario K1N 6N5, Canada}
\author{Graeme Bart}
\affiliation{Department of Physics, University of Ottawa, Ottawa, Ontario K1N 6N5, Canada}
\author{Andrew Parks}
\affiliation{Wyant College of Optical Sciences, University of Arizona, Tucson, Arizona, 85721, USA}
\author{Mohamad Toutounji}
\affiliation{College of Science, Department of Chemistry, UAE University, Al-Ain, UAE}
\author{Giulio Vampa}
\affiliation{Joint Attosecond Science Laboratory, National Research Council of Canada and University of Ottawa, 100 Sussex Drive, Ottawa,
Ontario K1A 0R6, Canada}
\author{Thomas Brabec}
\email{Thomas.brabec@uottawa.ca}
\affiliation{Department of Physics, University of Ottawa, Ottawa, Ontario K1N 6N5, Canada}
\author{Lu Wang (汪璐) }
\email{lu.wangTHz@outlook.com}
\affiliation{Department of Physics, University of Ottawa, Ottawa, Ontario K1N 6N5, Canada}

\begin{abstract}
\noindent 
Dephasing is the loss of phase coherence due to the interaction of an electron with the environment. The most common approach 
to model dephasing in light-matter interaction is the relaxation time approximation. Surprisingly, its use in intense laser physics
results in a pronounced failure, because ionization {is highly overestimated.} 
Here, this shortcoming is corrected by 
developing a strong field model in which the many-body environment is represented by a heat bath. Our model reveals that ionization
enhancement and suppression by several orders of magnitude are still possible, however only in more extreme parameter regimes. Our approach allows the integration of many-body physics into intense laser dynamics with minimal computational and mathematical complexity, 
thus facilitating the identification of novel effects in strong-field physics and attosecond {science}. 
\end{abstract}

\maketitle
\section{Introduction}

\noindent
Strong laser-matter interaction is commonly modeled as a closed quantum system with a single active electron 
\cite{krausz09,goulielmakis22}. While this assumption is well justified for atomic gases, its validity is not so clear for denser
materials, such as liquids and solids. A full many-body treatment of the non-perturbative dynamics of all electrons and nuclei is
prohibitively difficult. Therefore, it is more practical to model dense materials as a single active electron within an open quantum
system, where many-body effects are accounted for by interactions with the environment \cite{haug09, may23}. Due to its simplicity, 
the environment in intense laser-driven solids is mostly modeled in the relaxation time approximation
\cite{vampa2014theoretical,du22}, where the effect of many-body dynamics is replaced by a dephasing time $T_2$
\cite{witzel2010electron,paz1993reduction,yu2003qubit}. Dephasing represents the destruction of the coherence between
different one-electron eigenstates of the material, as a result of many-body collisions.

\begin{figure}[H]
\centering
\includegraphics[width=1\linewidth]{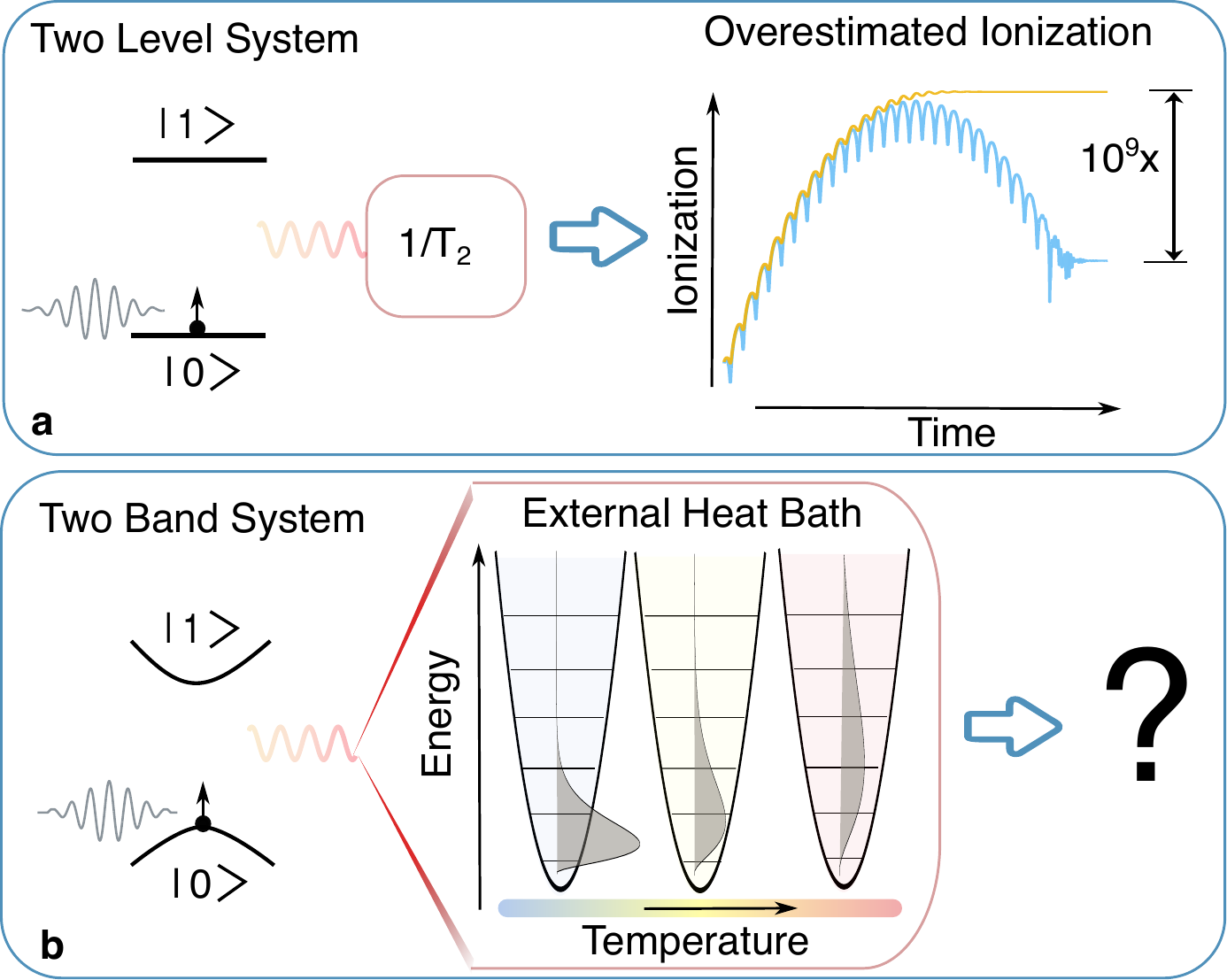}
\caption{Illustration of under-resonantly driven, open two-level/band systems. Panel \textbf{a} presents the two-level system (band gap
$E_g = 3.51$ eV) described by the relaxation time approximation. On the right-hand side ionization with ($T_2=8$\,fs, yellow curve) and
without dephasing ($T_2 = \infty$, blue curve) is compared. A moderate electric field strength $\text{E}_0=5\times 10^8\text{V}/\text{m}$ 
with photon energy $\sim 0.39$\, eV ($\lambda_0=3.2$\textmu m) is chosen. See Supplement Fig.S3 for details. Panel \textbf{b} shows the 
two-band system coupled to a heat bath described via the spin-boson model. The heat bath is modeled using boson harmonic oscillator modes. 
As the temperature rises, boson modes with higher energies are engaged (gray curves). } 
\label{fig:illustration}
\end{figure}

However, a simple calculation for an under-resonantly driven two-level system reveals questionable features of the relaxation time
approximation \cite{mcdonald2017strong}. Throughout this paper, we refer to ionization as the laser-induced excitation of an electron 
from the valence $\ket{0}$ to the conduction $\ket{1}$ band, as determined by conventional optical field ionization theory
\cite{keldysh1964ionization}. In Fig.
\ref{fig:illustration}\textbf{a} the ionization dynamics with dephasing described via the relaxation time approximation (yellow) and
without dephasing (blue) are compared. It can be seen that the relaxation time approximation predicts nine orders of magnitude ionization
enhancement at very moderate electric field strength. This is clearly unphysical {according to experimental measurements}. Though this overestimation of ionization may be mitigated by introducing a time-dependent relaxation time \cite{kruchinin2019non,cai2020quantum}, the underlying issue is still not resolved.  We term the ionization enhancement caused by dephasing as \textit{dephasing ionization}. 

In short, dephasing ionization happens when the phase relation between the laser and the two-level 
system is disturbed. Therefore, the laser-driven virtual population of the excited state is transformed into real excitation i.e. 
dephasing ionization. {Here, the virtual excitation refers to the population that disappears after the laser pulse.} The apparent shortcomings of the relaxation time approximation leave a gap between more complex and computationally 
demanding many-body approaches and oversimplified dephasing models commonly used in intense light-matter interaction. 
Furthermore, ionization is the first step in all strong field processes, such as material machining \cite{gattass08,yanik04,farsari09},
petahertz electronics \cite{schiffrin13,boolakee22}, electron acceleration from nano emitters \cite{chlouba23}, and attosecond spectroscopy 
in atoms, molecules and solids \cite{goulielmakis22,korolev2024unveiling}. Due to the importance of ionization, a deeper understanding of
dephasing ionization is essential. 

As such, a more sophisticated model is needed that ideally maintains most of the simplicity and wide applicability of the relaxation time
approximation. We borrow inspiration from the field of open quantum systems and adopt one of its key achievements, the spin-boson model, which typically serves as a minimal model to describe the quantum dynamics of an electron under the influence of the 
environment \cite{segal2005spin,leggett1987dynamics,thorwart2004dynamics}. Here, the spin-boson model is integrated into the 
semiconductor Bloch equations governing intense laser solid-state physics. The electron dynamics is represented by a single electron-hole, 
two-band model which is linearly coupled to its environment via bosonic harmonic oscillator modes, see Fig. 
\ref{fig:illustration}\textbf{b}. The so-called strong field spin-boson (SFSB) model allows for a closed-form solution of the electron 
dynamics in an environment and in the presence of an intense laser. We refer to the environment as a heat bath in the rest of the paper. 

The SFSB fixes the pathological ionization behavior displayed by the relaxation time approximation. Nevertheless, numerical analysis of
the SFSB equation reveals that ionization enhancement of up to a few orders of magnitude is still possible, but only at high temperatures. 
Interestingly, in the opposite low-temperature limit the heat bath can suppress ionization by up to a few orders of magnitude, which we 
term as \textit{dephasing suppressed ionization}. This occurs when the electron and heat bath interact strongly. 

The SFSB model provides a distinctive approach to uncovering the physics of complex many-body systems with minimal computational and
mathematical complexity. The predictive power of the SFSB approach can be progressively refined through either more detailed models or by 
fine-tuning the heat bath response through comparison with experiments. We anticipate that the SFSB framework will facilitate the 
discovery of new phenomena in strong-field physics and attosecond {science}.

\section{Theory}
Our analysis starts with a single electron two-band system coupled to a bosonic heat bath via a linear interaction term, \cite{lee2012accuracy,lambert2019modelling}
\begin{align}\label{eq:spin_boson}
   & H=-\frac{1}{2} \mathcal{E}(\bm{K}_t,t) \sigma_z+ \frac{1}{2}\hbar\Omega(\bm{K}_t,t) \sigma_x+\sum_q \hbar\omega_q 
   b_q^{\dagger} b_q \nonumber\\
   &+{\sigma_z} \sum_q g_q\left(b_q+b_q^{\dagger}\right) \mathrm{.} 
\end{align}
Here, $\mathbf{E}(t)$ is the laser electric field, the vector potential is defined by $-\partial_t \mathbf{A} = \mathbf{E}$, 
and $\bm{K}_t = \bm{K} + e \bm{A}(t) /\hbar$. The canonical momentum $\bm{K}$ belongs to the shifted Brillouin zone $\overline{\mathrm{BZ}}$. 
Further, $\Omega(\bm{{K}}_t,t) = 
(2 e / \hbar) \mathbf{d}(\bm{K}_t,t) \mathbf{ E}(t)$ is a generalized Rabi frequency, $e>0$ is the elementary charge and $\hbar$ is the
Planck constant; $\mathbf{d}(\bm{K}_t,t)$ and $\mathcal{E}(\bm{K}_t,t)$ represent transition dipole and bandgap between conduction 
$\ket{1}$ and valence $\ket{0}$ band, respectively. The time dependence of these quantities arises from the moving momentum frame. 
The Pauli matrices are denoted by $\sigma_j$ ($j =x,y,z$). Finally, $\omega_q$, $\hat{b}^\dagger_q, \hat{b}_q$, and $g_q$ are the 
harmonic oscillator frequency, creation, and annihilation operators, and the coupling coefficient of a mode with momentum $\mathbf{q}$,
respectively. 

{The coupling term between the heat bath and the two-band system appears exclusively in the diagonal terms of the Hamiltonian. Thus, it accounts only for dephasing, and not directly for heat-bath driven transitions between bands, i.e the off-diagonal terms. Nevertheless, due to the coupling of laser and heat
bath driven dynamics \mbox{\cite{vampa2014theoretical,wang2024tabletop}}, dephasing does influence the overall ionization. In the high-temperature limit, multi-boson transitions between valence and conduction band could become relevant but are ignored here.}

The Hamiltonian shown in Eq.(\ref{eq:spin_boson}) can be further simplified. First, we perform a polaron transformation that diagonalizes the 
laser-free Hamiltonian \cite{mahan13}. This is followed by a change to the interaction picture, which results in
\begin{align}
H_I=-\frac{\mathcal{E}(\bm{K}_t,t)}{2} \sigma_z+\frac{1}{2} \hbar \Omega(\bm{K}_t,t) \left(\sigma_{+} D^{\dagger^2}+\sigma_{-} D^2\right).
\label{eq:spin_boson_after_p}
\end{align}
For a detailed derivation, see Supplementary Material, Section I.  
Here, $\sigma_{+}=\left(\sigma_x+i \sigma_y\right) / 2 \text { and } \sigma_{-}=\left(\sigma_x-i \sigma_y\right) / 2$. The interactions
with laser and heat bath are now described by a single term, with the shift operator defined as $D=\exp \left\{-\sum_q 
{g_q} \left[b_q^{\dagger}(t)-b_q(t)\right]/({h \omega_q})\right\}$. 

The evolution of the density matrix is determined by the integration of the Liouville–Von Neumann equation with the Hamiltonian shown in Eq.(\ref{eq:spin_boson_after_p}). Initially, the valence band is fully occupied, the conduction band is empty, and the heat bath is in 
thermal equilibrium. A closed-form solution is obtained by using a Dyson expansion up to {the} second order. As we are only interested 
in the two-band system dynamics, the heat bath degrees of freedom are traced out (see Supplementary Material Sections II-III for 
details) \cite{lee2012accuracy,wurger1998strong,nicolin2011non,morreau2019phonon,bundgaard2021non,liu2014reduced,thorpe2023high,
lambert2019modelling,meier1999non,paz1993reduction,rouse2022analytic}. We found that the dominant contribution to ionization is 
contained in the second order expansion term \cite{thorpe2023high} from which the conduction band population follows as 
\begin{align}
&n_c({\bm{K}},t) = \frac{1}{2} \operatorname{Re}\left\{\int_{-\infty}^t \int_{-\infty}^{t_1} \Omega^*\left(\bm{K}_{t_1},t_1\right) 
\Omega\left(\bm{K}_{t_2},t_2 \right) \right.\nonumber\\
    &\biggl.\times\exp \left[iS(t_1,t_2)+C\left(t_1-t_2\right)\right] d t_1 d t_2\biggr\}\label{eq:rho_kt},\\
 &  n_c(t) = \int_{{\overline{\mathrm{BZ}}}} \! n_c({\bm{K}},t) d{\bm{K}},\label{eq:rho_final} 
\end{align}
where the action $S(t_1,t_2)=\int_{t_2}^{t_1}d\tau\,\mathcal{E}_s(\bm{K}_\tau,\tau)/\hbar$, and $\mathcal{E}_s(\bm{K}_{\tau},\tau) =
\sqrt{\mathcal{E}(\bm{K}_{\tau},\tau)^2+\hbar^2\Omega(\bm{K}_{\tau},\tau)^2}$ is the bandgap shifted by the dynamic Stark effect \cite{thorpe2023high,toth2023role,sussman2011five}. One can see from Eq.(\ref{eq:rho_kt}) that the heat bath influences are exclusively included in 
the correlation function 
\begin{align}
& C\left(t_1-t_2\right) \approx \int_{-\infty}^{\infty} J(\omega)\biggl\{ i \sin \left[\omega\left(t_1-t_2\right)\right]\biggr.
\nonumber\\
&\left.-\left\{1-\cos \left[\omega\left(t_1-t_2\right)\right]\right\} \operatorname{coth} \left(\frac{\hbar \omega}{2k_BT}\right) \right\}
d \omega ,\label{eq:J_t}
\end{align}
where $k_B$ is the Boltzmann constant. The temperature $T$ dependence in Eq.(\ref{eq:J_t}) is contained only in the $\coth$ term. The 
$g_q$ related terms in Eqs.~(\ref{eq:spin_boson},\ref{eq:spin_boson_after_p}) are replaced by a spectral density $J(\omega)$ 
through a transition from discrete to continuous modes. The spectral density depends on two parameters: coupling strength $j_o$, 
and cutoff frequency $\omega_c$. There exists a wealth of different models for the 
spectral density $J(\omega)$, such as the Debye \cite{liu2014reduced}, Ohmic \cite{paz1993reduction}, Under-Damped Brownian
\cite{lambert2019modelling,meier1999non}, Gaussian \cite{rouse2022analytic}, and Shifted-Gaussian models, the definition of which can 
be found in the Supplementary Material, Sec. IV. 

The relaxation time approximation is recovered for the Debye bath in the high $T$-limit, $C(t_1-t_2) \rightarrow -(t_1-t_2)/T_2$ with 
$T_2=\hbar/(2\pi k_B Tj_o)$, as outlined in the Supplementary Material, Section IV.A. By contrast, the high $T$-limits of the other heat
bath models do not exhibit a linear time dependence in the exponent. 

In the context of strong laser solid interaction, the temperature $T$ refers to the local electron or ion temperature. Our approach 
presents an approximation, as the system, its dependence on laser pulse duration, is not always in thermal equilibrium. This can be 
analyzed via the well-established two-temperature model, where electrons are first heated by the laser, and then the energy is transferred to the
lattice, raising its temperature. Material damage or melting is typically determined by the lattice temperature. For dielectrics, 
damage occurs around a few thousand K, even though the electron temperature can be much higher, reaching up to $10^5\,\text{K}$
\cite{chen2006semiclassical,mozafarifard2023two,carpene2006ultrafast}. While our approach can be extended to describe non-equilibrium 
heat baths, this would go beyond the limit of an initial investigation.

The cutoff frequency $\omega_c$ falls within the Terahertz to the far-infrared range for phonons, and spans the far-infrared 
to the mid-infrared range for collective electronic excitations, such as excitons and plasmons. The coupling strength $j_o$ is a
dimensionless parameter ranging
from $10^{-3}$ to multiples of unity 
\cite{lambert2019modelling,lee2012accuracy,yamamoto2022heat,anto2021strong,franchini2021polarons,magazzu2018probing}. For phonons, 
$j_o < 1$ in III-V semiconductors, whereas $j_o > 1$ in more polar II-VI compounds \cite{haug09}. {Strong electron-phonon coupling  $j_o >1$ typically occurs in very polar materials \mbox{\cite{franchini2021polarons,devreese2009frohlich}} such as bi-layer graphene \mbox{\cite{chen2024strong}}, single-layer InSe \mbox{\cite{lugovskoi2019strong}} and superconductors \mbox{\cite{wu2024ultrafast,errea2020quantum}}}. For collective electronic excitations, the coupling strength 
depends on the electron density \cite{caruso2016theory}. For electron densities above $10^{20}$cm$^{-3}$ and for $\hbar\omega_c \sim 1\,$eV
the plasmon coupling strength can become comparable to and even exceed the phonon coupling strength. 

%
\section{Results}
We have selected zinc oxide (ZnO), a representative and widely studied semiconductor. The crystal momentum $\bm{k}$ dependence in the entire 3D Brillouin zone is considered for the two-band system. Material parameters are derived from ab initio calculations \cite{goano2007electronic,vampa2015semiclassical,vampa2015all} (see Supplementary Material Section V, Table I). We find that both 3D and 1D calculations along the $\Gamma$-M direction yield similar results in terms of relative heat bath induced ionization changes, both quantitatively and qualitatively (see Supplementary Material Fig.S4). Therefore, for computational efficiency, we focus on the 1D Brillouin zone along the $\Gamma$-M direction throughout the following calculations.

A driving laser with the center 
wavelength $\lambda_0=3.2\,$\textmu m is selected. The center frequency is defined as $\omega_0=2\pi c/\lambda_0\approx 2\pi \times 
10^{14}$\,Hz  ($\hbar\omega_0 \sim 0.39$\, eV) with $c$ the vacuum light velocity. The energy of the laser photons is much lower than the
resonance energy of ZnO (with a band gap of $\mathcal{E}_g = 3.51$ eV), meaning that at least 9 photons are required to excite an electron from 
the valence band to the conduction band. 
We choose a linearly polarized electric field defined as 
$\mathbf{E}=\text{E}_x=\text{E}_0\exp{\left(-t^2/\tau^2\right)}\cos{(\omega_0t)}$, where $\tau=20\,\text{fs}$. The electric field strength 
$\mathrm{E}_0=1.5\times10^9\,\text{V}/\text{m}$ is well below the single pulse damage threshold of ZnO \cite{dufft2009femtosecond}. 
These parameter values are used throughout the paper unless otherwise stated.%
\begin{figure}[h]
\centering
\includegraphics[width=1\linewidth]{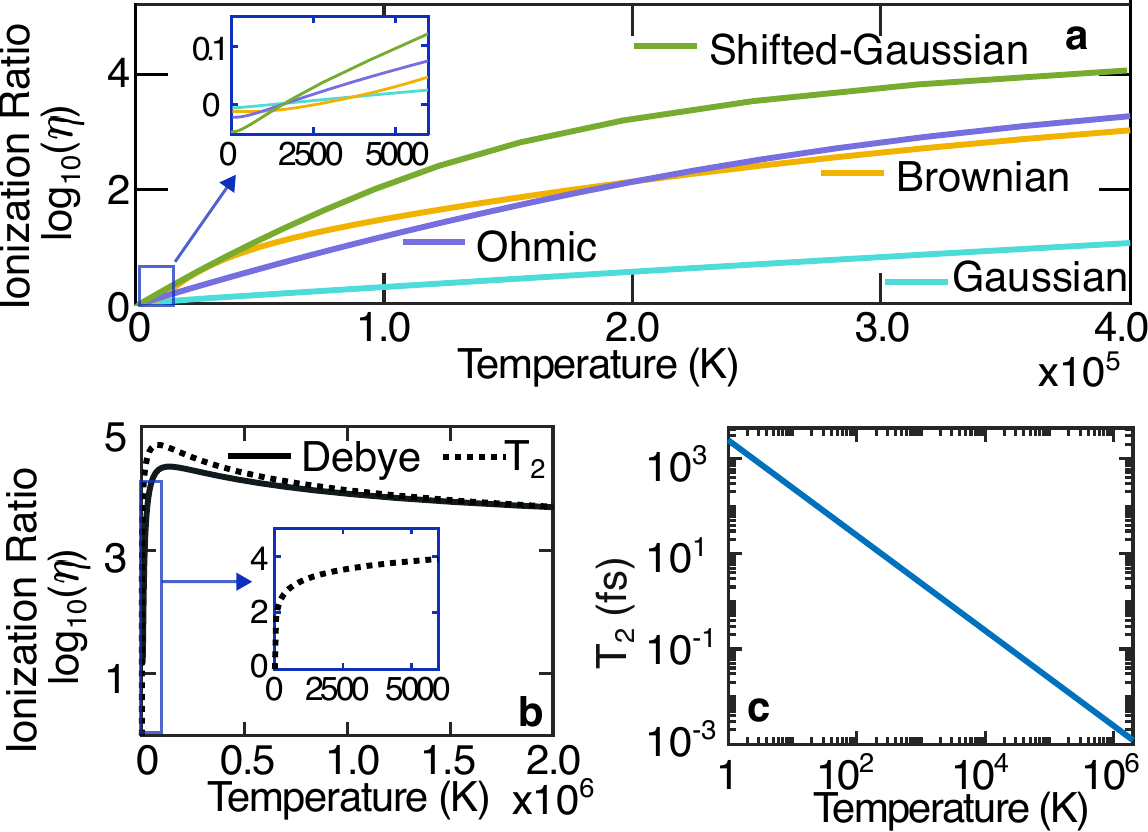}
\caption{ Panel \textbf{a} presents the ionization ratio versus temperature $T$ for various heat baths. Panel \textbf{b} shows the ionization ratio for the Debye heat bath and relaxation time 
approximation versus $T$. The insets in \textbf{a} and \textbf{b} show details in the low $T$ regime. The relaxation time $T_2 = \hbar/(2\pi k_B T j_o)$  obtained from the Debye spectral density, is plotted in \textbf{c} as a function of $T$. The heat bath parameters are 
$\omega_c=0.1\omega_0$, $j_o=0.1$. }
\label{fig:T2_limit}
\end{figure}

The change of ionization due to the heat bath is characterized by calculating the ionization ratio with and without the heat bath, 
\begin{equation}
\eta=\left.\frac{n_c(j_o\neq0)}{n_c(j_o=0)}\right\vert_{t=\infty},
\end{equation}
where $n_c(t)$ is defined in Eq.(\ref{eq:rho_final}). 

In Fig.\ref{fig:T2_limit}\textbf{a}, the ionization ratio $\text{log}_{10}(\eta)$ is plotted versus $T$ for Ohmic, Under-Damped Brownian,
Gaussian, and Shift-Gaussian spectral densities, all of which follow a similar trend and yield comparable results. Thus, without loss of 
generality, we have chosen the Ohmic spectral density throughout the entire numerical analysis. The ionization ratio is plotted in $\text{log}_{10}$ scale, where the positive (negative) numbers of $\text{log}_{10}(\eta)$ correspond to the order of magnitude of enhancement (suppression) of ionization. Figure \ref{fig:T2_limit}\textbf{b} shows
that the Debye spectral density converges to the relaxation time approximation at very high temperatures. The temperature dependence of
$T_2$, obtained from the Debye spectral density in the high $T$ limit above, is presented in Fig. \ref{fig:T2_limit}\textbf{c}. Both Debye
and relaxation time approximation show an unrealistic rise of $\eta$ at low $T$ and therefore do not represent realistic heat bath models. 
This is to be expected, due to the unphysically long high-frequency tail of the Debye spectral density 
\cite{devreese2009frohlich,mishchenko2000diagrammatic}. Finally, by comparing the zoomed-in sections of Figs. \ref{fig:T2_limit}\textbf{a}
and \textbf{b}, one can see that the relaxation time approximation substantially overestimates ionization at low temperatures, while
all the other heat baths in \textbf{a} show negligible changes in ionization, as detected by experiments. 

\begin{figure}[H]
\centering
\includegraphics[width=1\linewidth]{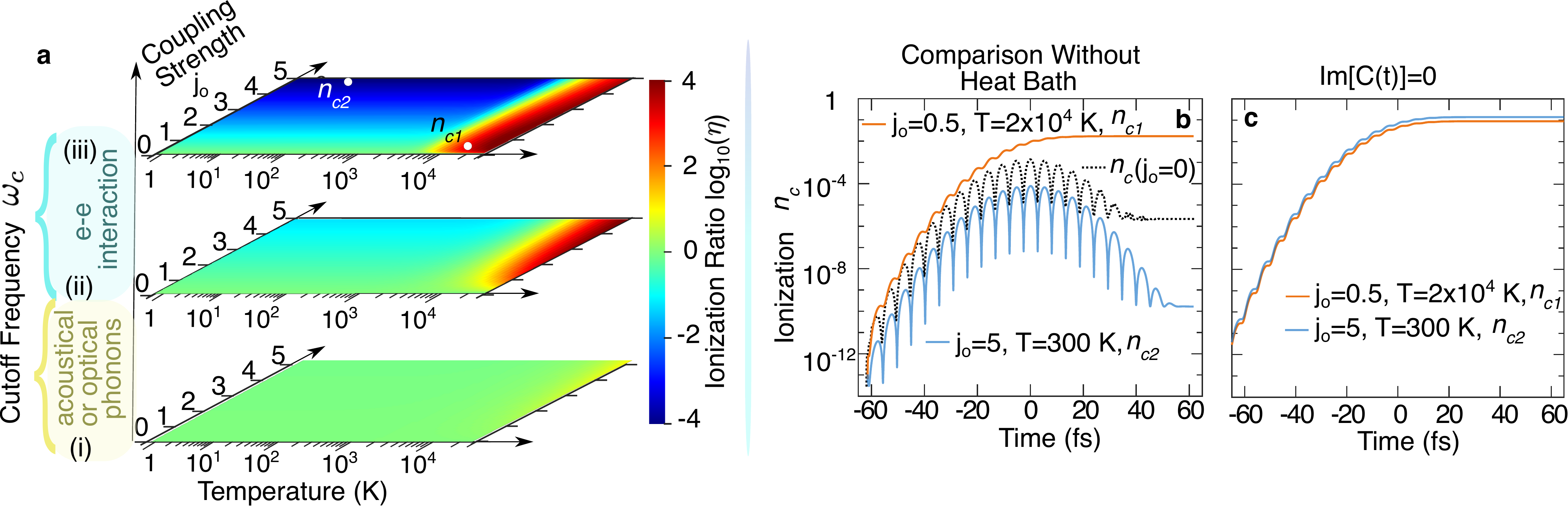}
\caption{Panel \textbf{a} shows ionization ratio $\mathrm{log}_{10}(\eta)$ as a function of local temperature $T\in[1,\,3\times10^4]$\,K and coupling coefficient $j_o\in[0,5]$. The three panels represent different cutoff frequencies, (i) $\omega_c=0.01\omega_0$, (ii) 
$\omega_c=0.1\omega_0$ and (iii) $\omega_c=2.1\omega_0$. \textbf{b} Ionization versus time for two data points $n_{c1}$ and $n_{c2}$ 
in panel (iii) of \textbf{a}; black dotted curve shows ionization in the absence of a heat bath. \textbf{c} same as plots for $n_{c1}$ and $n_{c2}$ in \textbf{b} only with setting the imaginary part of the heat bath response $C(t)$ [defined in Eq.(\ref{eq:J_t})] to zero. }\label{fig:three_bath}
\end{figure}

In Fig.\ref{fig:three_bath}\textbf{a}, $\mathrm{log}_{10}(\eta)$ is scanned over a wide range of $T$ and $j_o$ for three representative
values of $\omega_c$, referring to various collective lattice or electron excitations; (i) $\omega_c = 0.01\omega_0$, (ii) $\omega_c = 0.1\omega_0 $, and (iii) $\omega_c = 2.1 \omega_0$. Acoustic and optical phonons span the range from (i) to (ii), whereas collective electronic excitations span the range from (ii) to (iii) with excitons for ZnO around (ii) \cite{fiedler2020correlative} and
laser-excited plasmons in the spectral range around (iii) and above \cite{koch2023heavily,kesim2014plasmonic}. Although electrons are fermions, their collective excitations can, to a good approximation, be treated as bosons \cite{caruso2016theory,lambert2019modelling,devreese2009frohlich}. As such, they can be 
directly modeled via the spin-boson Hamiltonian shown in Eq.(\ref{eq:spin_boson}). While all $T$ and $\omega_c$ ranges can be realized in intense 
laser-driven ZnO, the shown $j_o$-dependence is not ZnO specific. We explore the typical range of $j_o$ defined above.

\onecolumngrid
\,
\FloatBarrier
\twocolumngrid  

The panel (i) of Fig.\ref{fig:three_bath}\textbf{a} represents acoustic phonons, suggesting the heat bath has little effect on ionization. The influence
of the heat bath increases with $\omega_c$, as seen in (ii) and (iii), which embody optical phonons and electronic excitations. In the 
high-$T$ limit, ionization is increased by several orders of magnitude, which we call dephasing ionization. On the other hand, at moderate $T$ with strong 
coupling ($j_o > 1$), ionization is suppressed by multiple orders of magnitude, which we have termed dephasing suppressed ionization. These two limits are represented by data points $n_{c1}$, $n_{c2}$ in panel (iii) for which, the temporal evolution of ionization is
plotted in Fig. \ref{fig:three_bath}\textbf{b}. The black dotted curve represents ionization in the absence of a heat bath $n_{c}(j_o=0)$.

The increase and decrease of ionization can be explained by the real and imaginary parts of the correlation function $C(t)$. With a given $\omega_c$, at extremely 
high temperatures, the correlation function approaches a delta function (instantaneous) in time, leading to the Markovian limit 
\cite{hofer2017markovian}. In this limit, the real part of the correlation function dominates, and one may neglect the imaginary 
contribution. This is why the relaxation time approximation using $T_2$ as a purely real number remains a valid approximation at high
temperatures. On the other hand, at low temperatures, the correlation function is non-Markovian with a wider distribution in time. 
In this case, the phase of the correlation function acts as a dynamic addition to the bandgap, increasing the original material bandgap, and thereby resulting in dephasing suppressed ionization. The importance of the heat bath phase becomes clear from a comparison of
\ref{fig:three_bath}\textbf{b} and \ref{fig:three_bath}\textbf{c}. In Fig.\ref{fig:three_bath}\textbf{c} the imaginary part of the
$C(t)$ is set to zero, as a result of which ionization at $T=300$\,K changes from suppression into enhancement. 

\begin{figure}[b]
\FloatBarrier
\centering
\includegraphics[width=1\linewidth]{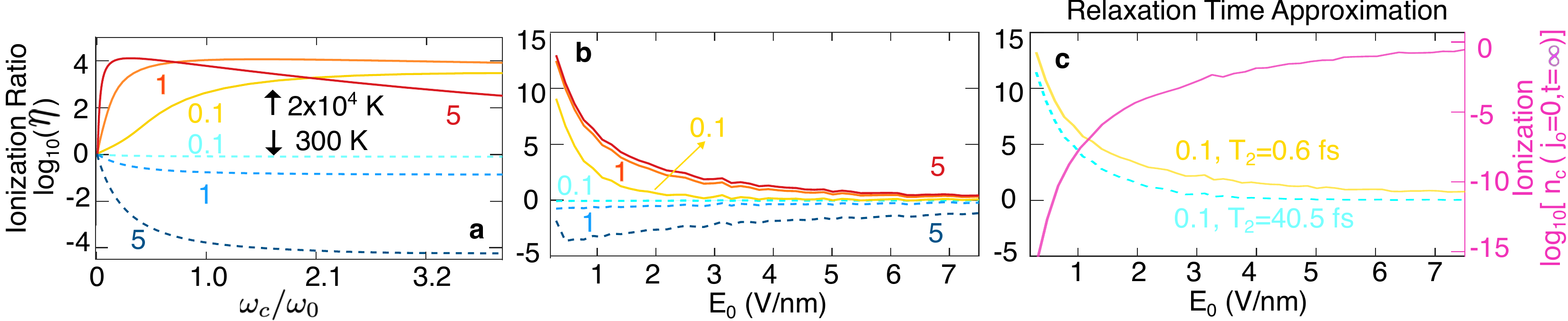}
\caption{Ionization ratio as a function of cutoff frequency $\omega_c$ (panel \textbf{a}) and of peak electric field strength $\mathrm{E}_0$ (panels \textbf{b,c}) are presented. We have chosen different values of $j_o\in\{0.1,1,5\}$ denoted by different colors beside each curve. The cold-colored dashed curves are for $T=300$K; warm-colored full curves refer to $T = 2\times 10^4$ K.  In \textbf{a}, the ionization without the heat bath is $n_c(j_o=0,t=\infty)=2\times10^{-6}$. Panels \textbf{b,c} are calculated by $\omega_c=0.4\omega_0$. The relaxation time used in panel \textbf{c} is calculated by $T_2=\hbar/2\pi k_Bj_o T$. The pink curve plotted on the right $y$ axis shows the ionization $n_c(j_o=0,t=\infty)$ in the absence of the heat bath.}\label{fig:3}
\end{figure}

To gain further insight into the parameter dependence, ionization ratios are presented as a function of cutoff frequency $\omega_c$ in 
Fig.\ref{fig:3}\textbf{a} and of peak electric field strengths $\mathrm{E}_0$  in Figs.\ref{fig:3}\textbf{b,c}. We have chosen two 
different temperatures: 300\,K (represented by the cold color dashed curves) and $2 \times 10^4$\,K (represented by the warm-colored 
curves). The curves are color-coded to indicate different coupling strengths $j_o$, with the values of $j_o$ denoted in the same 
color. 

\onecolumngrid
\,\vspace{-0.5cm}%
\twocolumngrid
Figure \ref{fig:3}\textbf{a} confirms that dephasing ionization only occurs at high temperatures, while dephasing suppression ionization happens exclusively at low temperatures.  Figure \ref{fig:3}\textbf{b} indicates that the heat bath only plays a role at moderate electric field strengths. This can be explained by the multi-photon and tunneling ionization channels. When the electric field is strong, the Keldysh parameter $\gamma=\omega_0\sqrt{m^* \mathcal{E}_g}/(e\mathrm{E}_0)$ where $m^*$ is the effective mass and $\mathcal{E}_g$ is the band gap energy, becomes smaller than 1, suggesting the tunneling effects dominate \cite{keldysh2024ionization}. With our choice of parameters, $\gamma=1$ corresponds to $E_0\approx1.2\,$V/nm. Since tunneling ($\gamma<1$) occurs much more rapidly than multiphoton absorption \cite{landsman2014ultrafast,klaiber2015tunneling}, the heat bath cannot follow the ionization process and thus has negligible influence at large $\text{E}_0$. In addition, while  
optical field ionization scales exponentially with $\text{E}_0$, dephasing ionization scales proportional to the laser intensity 
\cite{mcdonald2017strong}. As a result, the relative importance of dephasing ionization drops for increasing laser fields.  The multi-photon ionization ($\gamma>1$) develops over an optical cycle and thus is more sensitive to the non-Markovian heat bath, making it more sensitive to heat bath influences.

In order to relate the relative ionization changes to absolute values, ionization in the absence of the heat bath $n_c(j_o=0,t=\infty)$, 
is shown as a function of $E_0$ in Fig.\ref{fig:3}\textbf{c}. At the highest field strength, ionization is approaching saturation. 
Moreover, the ionization ratio calculated via the relaxation time approximation is also presented. Comparing Figs.\ref{fig:3}\textbf{b,c}, 
one can see that the relaxation time approximation predicts orders of magnitude higher ionization compared to that predicted by our model.

\section{Discussion}

\noindent
So far, we have seen that the environment can modify ionization by orders of magnitude in the extreme limits of high $T$ or strong
coupling $j_o$. The environment in intense laser-solid interaction is difficult to control. There are various ways in which the 
environment can be engineered for more controlled experiments on dephasing and dephasing suppressed ionization. 

First, light modes in high-quality micro and nano-cavities can be controlled to vary from sub-poissonian, super-poissonian, poissonian, 
and squeezed vacuum to thermal distributions; from weak to strong coupling with electrons \cite{wei2024emission,najer2019gated}. 
As such, they can serve as an artificial, strongly coupled environment in which the modification of strong field processes by ionization 
can be investigated.

Second, collective electron oscillations can be created in tailor-made experiments. The conduction band can be populated by doping 
semiconductors, or with a pump pulse in a pump-probe experiment. Ionization changes are probed with a second pulse or with
transient absorption spectroscopy. As some of the effects observed here depend on strong coupling with the environment, control 
of the coupling strength is important. Coupling strength increases when going from bulk to 2D and 3D nano-scale materials, such as in nano-resonators and -cavities \cite{di2024toward,akerboom2024free}.

The possibility of engineering ionization has potential practical impacts. First, dephasing ionization increases ionization and thus, 
allows material micro-machining and -modification at lower laser intensities. This could be instrumental in generating highly charged ion states in high-density plasmas with lower pump pulse energy, contributing to the improvement of table-top X-ray sources. Second, the transition between perturbative nonlinear optics and 
strong field physics is marked by the onset of ionization. Dephasing suppressed ionization shifts this onset and {permits probing dynamics in materials under excitation conditions previously unattainable}.

\section{Conclusion}
\noindent 
The relaxation time approximation is frequently used in intense laser field physics to account for the many-body coupling between a single electron and its environment, which consists of lattice, impurities, and remaining electrons. This work aimed to understand the failure of the relaxation time approximation and to correctly describe ionization in an open quantum system. Ionization in the presence of the relaxation time approximation is enhanced by orders of magnitude over a wide range of parameters, which is termed dephasing ionization. 

To decide whether dephasing ionization holds physical significance or is simply a failure of the relaxation time approximation, we have 
developed a more comprehensive model that captures more physics and still retains much of the simplicity of the relaxation 
time approximation. 

Our results confirmed that ionization enhancement through dephasing ionization still persists, but only in fairly extreme
parameter ranges. Very little enhancement is found for acoustical phonon frequencies. For optical phonons and collective electronic
excitations dephasing ionization becomes dominant in the limit of high temperatures. Our analysis has also revealed the possibility that a heat bath can suppress ionization by orders of magnitude, which we have named dephasing suppressed ionization.

We presented a novel framework here to model intense laser many-body processes in a low-cost, semi-phenomenological way. Future research will entail finding optimum heat baths and complementing the current framework with more physics. In addition, a simple fermionic heat bath. Though the SFSB presents a good approximation to a large class of collective excitations of 
electrons and lattice, it does not account for electron-electron scattering which requires an extended approach with a fermionic heat bath 
\cite{michishita20}. Besides, heat bath parameters, such as material temperature, change during intense laser interaction. As such, the ionization dynamics investigated here present only approximate snapshots. For a full treatment of laser material interaction, a dynamically evolving heat bath will have to be considered.

\section{Acknowledgements}
L. W. would like to thank Fluffy B. for being the motivation to complete this project.

\end{CJK*}

\bibliographystyle{apsrev4-1} 
\bibliography{apssamp_1} %

\end{document}